\documentstyle[amssymb,aps]{revtex}

\draft

\begin{document}

\tighten

\twocolumn[\hsize\textwidth\columnwidth\hsize\csname@twocolumnfalse\endcsname

\title{Macroscopic quantum coherence in antiferromagnetic molecular magnets}

\author{Hui Hu$^{a}$, Rong L\"{u}$^{b}$, Jia-Ling Zhu$^{a,b}$, Jia-Jiong Xiong$^{a}$}
\address{$^a$Department of Physics, Tsinghua University, Beijing 100084, China\\
$^b$Center for Advanced Study, Tsinghua University, Beijing 100084, China}

\date{\today}
\maketitle

\begin{abstract}
The macroscopic quantum coherence in a biaxial antiferromagnetic molecular 
magnet in the presence of magnetic field acting parallel to its hard anisotropy 
axis is studied within the two-sublattice model. On the basis of instanton 
technique in the spin-coherent-state path-integral representation, both the rigorous 
Wentzel-Kramers-Brillouin exponent and preexponential factor for the ground-state tunnel 
splitting are obtained. We find that the quantum fluctuations around the classical paths 
can not only induce a new quantum phase previously reported by Chiolero and Loss
(Phys. Rev. Lett. 80, 169 (1998)) , but also have great influnence on the intensity of 
the ground-state tunnel splitting. Those features clearly have no analogue in the ferromagnetic 
molecular magnets. We suggest that they may be the universal behaviors in all antiferromagnetic 
molecular magnets. The analytical results are complemented by exact diagonalization calculation.
\end{abstract}


]

\section{Introduction}

In recent years, owing mainly to the rapid advances both in new technologies
of miniaturization and in highly sensitive SQUID magnetometry, there have
been considerable theoretical and experimental studies carried out on the
nanometer-scale magnets \cite{Mn12Ac,Fe8} which have been identified as
candidates for the observation of macroscopic quantum phenomena (MQP) \cite
{gunther,chud98} such as the tunneling of the spin out of metastable
potential minimum through the classically impenetrable barrier to a stable
one, i.e., macroscopic quantum tunneling (MQT), or, more strikingly,
macroscopic quantum coherence (MQC), where the spin coherently oscillates
between energetically degenerate easy directions over many periods. In the
semiclassical spin-coherent-state path-integral theory \cite{fradkin}, MQC
is connected with the presence of a topological term in the Euclidean action 
$S_E(\theta ,\phi )$ arising from the nonorthogonality of spin coherent
states, which is called Berry phase or the Wess-Zumino, Chern-Simons term:\ $%
iS\int\limits_{-\frac T2}^{+\frac T2}(1-\cos \theta )\dot{\phi}\left( \tau
\right) d\tau $, where $S$ is the whole spin of the system, $\left( \theta
,\phi \right) $ are polar and azimuthal spin angles, repectively.

One of the manifestations of MQC is the ground-state tunnel splitting of
magnetic systems. In the absence of an external magnetic field, it has been
theoretically demonstrated that the ground-state tunnel splitting is
completely suppressed to zero for the half-integer total spin ferromagnets
or antiferromagnets with biaxial crystal symmetry \cite{loss,delft},
resulting from the destructive interference of the Berry phase in the
Euclidean action between the symmetry-related tunneling paths connecting two
classically degenerate minima. Such destructive interference effect for
half-integer spins is known as the topological quenching. But for the
integer spins, the quantum interference between topologically different
tunneling paths is constructive, and therefore the ground-state tunnel
splitting is nonzero.

While such spin-parity effects are sometimes related to Kramers degeneracy,
they typically go beyond this theorem in rather unexpected ways. In the
presence of an external magnetic field, as pointed out by Garg \cite
{garg93,garg99}, the ground-state tunnel splitting can oscillate as a
function of the field which is applied along its hard anisotropy axis in
ferromagnets with biaxial crystal symmetry, and vanishes at certain values.
This prediction is confirmed in one recent experiment carried out by
Wernsdorfer and Sessoli \cite{wernsd}. They developed a new technique to
measure the very small tunnel splitting on the order of $10^{-8}$K in
ferromagnetic molecular Fe$_8$ clusters. Indeed, they observed a clear
oscillation of the tunnel splitting as a function of the magnetic field
applied along the hard anisotropy axis, which is direct evidence of the role
of the topological spin phase (Berry phase) in the spin dynamics of these
molecules. Although this field induced oscillation's behaviour is
investigated in great detail in ferromagnetic systems now \cite
{garg93,garg99,wernsd}, it is still less  understood in antiferromagnetic
molecular magnets such as Fe$_{10}$, Fe$_6$, V$_8$, and antiferromagnetic
ferritin \cite{golyshev,chiolero}. Golyshev and Popkov first studied MQC in
a uniaxial antiferromagnetic fine particle in the presence of magnetic field 
\cite{golyshev}, and found similar oscillation behavior. But they only
calculated the Wentzel-Kramers-Brillouin (WKB) exponent in weak field
approximation, and paid no attention to its preexponential factor. Later, in
1998, Chiolero and Loss considered the oscillation's properties of a
ringlike molecular magnet using an anisotropic nonliner $\sigma $ model
(NLsM) \cite{chiolero}. In addition to the usual topological spin phase
(Berry phase) term, they found a new quantum phase arising from fluctuations
which is never seen in ferromagnetic molecular magnets. It is really a
striking quantum property in antiferromagnetic molecular magnets.
Unfortunately, the fundamental physics of this novel quantum phase is less
explored so far.

In this paper, We would like to study the MQC of a biaxial symmetry
antiferromagnetic molecular magnet based on the two-sublattice model \cite
{barbara,chud95,sima}. By applying the instanton technique in the
spin-coherent-state path-integral representation \cite{garg92}, we obtain
the rigorous instanton solutions and calculate both the WKB exponent and
preexponential factor in the ground-state tunnel splitting. We will show
that the quantum fluctuations around the classical paths can not only induce
a new quantum phase previously reported by Chiolero and Loss \cite{chiolero}%
, but also have great influence on the intensity of the ground-state tunnel
splitting. Those features clearly have no analogue in the ferromagnetic
molecular magnets. We suggest that they may be universal behaviors in all
antiferromagnetic molecular magnets. Due to the instanton methods are
semiclassical in nature, i.e., valid in large spins and in continuum limit,
we perform exact diagonalization calculations and find that they agree well
with the analytical results.

\section{Instanton calculations for spin-coherent-state path integrals}

We consider the ring-like antiferromagnetic molecular magnets(i.e. Fe$_{10}$%
, Fe$_6$ and V$_8$) composed of $N=2n$ spins $s$ regularly spaced on a
circle lying in the $xy$-plane with an antiferromagnetic exchange
interaction between them \cite{chiolero,zvezdin}. In general, the
crystalline anisotropy at each site has biaxial symmetry. As usually for
antiferromagnets \cite{barbara,chud95,sima}, we decompose the local spins
into the two magnetic sublattices: $\vec{S}_1$ and $\vec{S}_2$ with the same
spin value $S=ns.$ Then, the molecular magnet in an external magnetic field $%
\vec{H}$ acting along its hard anisotropy axis can be described by a spin
Hamiltonian of the type 
\begin{eqnarray}
{\cal H} &=&j\vec{S}_1\cdot \vec{S}_2+\left( k_1\hat{S}_{1z}^2+k_2\hat{S}%
_{1y}^2-g\mu _BH\hat{S}_{1z}\right) +  \nonumber \\
&&\left( k_1\hat{S}_{2z}^2+k_2\hat{S}_{2y}^2-g\mu _BH\hat{S}_{2z}\right) ,
\label{totHami}
\end{eqnarray}
where $g$ is the land\'{e} factor, and $\mu _B$ is the Bohr magneton. $%
k_1>k_2>0$ are the crystalline anisotropy coefficients, and we take the
easy, medium, and hard axes as {\bf x}, {\bf y}, and {\bf z} respectively
for each sublattice. $j$ is the exchange energy. In accordance with
experimental results it will be assumed that $j\gg $ $k_1,k_2$ for the
strong antiferromagnetic coupling. Note that our two-sublattice
configuration is only valid for the magnetic field $H\leqslant H_a$. Here, $%
H_a=\frac{2jS}{g\mu _B}$ is the critical field at which the strong
antiferromagnetic exchange interaction $j\vec{S}_1\cdot \vec{S}_2$ is
comparable to the Zeeman term $g\mu _BH(\hat{S}_{1z}+\hat{S}_{2z})$.

In the semiclassical approach \cite{garg92}, in order to obtain the
ground-state tunnel splitting, one should compute the imaginary-time
propagator in the spin-coherent-state representation: 
\begin{eqnarray}
&&\left\langle \hat{n}_f\left| exp\left[ -{\cal H}T\right] \right| \hat{n}%
_i\right\rangle  \nonumber \\
&=&\int {\cal D}\Omega \exp \left( -S_E\right)  \nonumber \\
&=&\int {\cal D}\{\theta _1\}{\cal D}\{\theta _2\}{\cal D}\{\phi _1\}{\cal D}%
\{\phi _2\}\exp \left( -\int d\tau {\cal L}\right)
\end{eqnarray}
over all trajectories which connect the initial state $\mid \hat{n}_i\rangle 
$ to the final state $\mid \hat{n}_f\rangle $. Here $\theta _j$, $\phi _j$ $%
(j=1,2)$ are the polar and azimuthal angles of each sublattice spin vector,
the Lagrangian ${\cal L}$ include two parts \cite{golyshev}: 
\begin{equation}
{\cal L}_0=\sum\limits_{j=1,2}iS(1-\cos \theta _j)\dot{\phi}_j\left( \tau
\right)
\end{equation}
and 
\begin{eqnarray}
{\cal L}_1 &=&J\left[ 1+\cos \theta _1\cos \theta _2+\sin \theta _1\sin
\theta _2\cos \left( \phi _1-\phi _2\right) \right]  \nonumber \\
&&+\sum\limits_{j=1,2}\left( K_1\cos ^2\theta _j+K_2\sin ^2\theta _j\sin
^2\phi _j\right)  \nonumber \\
&&-\sum\limits_{j=1,2}g\mu _BSH\cos \theta _j,  \label{energy}
\end{eqnarray}
corresponding to the Berry phase term and the total Euclidean energy term $%
E(\theta _1,\phi _1,\theta _2,\phi _2)$. Here, we have introduced $%
K_1=k_1S^2 $, $K_2=k_2S^2$, and $J=jS^2$. All terms in (\ref{energy}) are of
apparent physical meaning. The first term is the exchange interaction
energy, the second term is magnetic anisotropy energy and the third term is
Zeeman energy. The dominant contribution to the imaginary-time propagator
comes from finite action solutions of the Euler-Lagrange equations of motion
(instatons), which can be expressed as 
\begin{eqnarray}
\frac{\delta S_E}{\delta \bar{\theta}_j} &=&0,  \label{Ea} \\
\frac{\delta S_E}{\delta \bar{\phi}_j} &=&0,  \label{Eb}
\end{eqnarray}
where $\bar{\theta}_j$, $\bar{\phi}_j$ $(j=1,2)$ denote the classical paths.

According to the instanton technique in the spin-coherent-state
path-integral representation \cite{garg92}, the instanton's contribution to
the tunnel splitting $\Delta $ (not including the geometric phase factor
generated by the Berry phase term in the Euclidean action) is given by 
\begin{equation}
\Delta =p_0\omega _p\left( \frac{S_{cl}}{2\pi }\right) ^{1/2}e^{-S_{cl}},
\end{equation}
where $\omega _p$ is the small-angle precession or oscillation frequency in
the well, and $S_{cl}$ is the classical action or the WKB exponent
determined by Eqs. (\ref{Ea}) and (\ref{Eb}). The preexponential factor $p_0$
originates from the quantum fluctuations around the classical paths, which
can be evaluated by expanding the Euclidean action to second order in the
small fluctuations.

\subsubsection{Wentzel-Kramers-Brillouin exponent}

In our case, only low-energy trajectories with almost antiparallel $\vec{S}%
_1 $ and $\vec{S}_2$ contribute the path integral. It is therefore, safe to
say that tunneling of $\vec{S}_2$ follows tunneling of $\vec{S}_1$ \cite
{chud95}. For that reason we can replace $\theta _2$ and $\phi _2$ by $\pi
-\theta _1\,+\varepsilon _\theta $ and $\pi +\phi _1-\varepsilon _\phi $
respectively (with $\left| \varepsilon _\theta \right| ,\left| \varepsilon
_\phi \right| \ll 1$) in ${\cal L}.$ In the new coordinates, the
imaginary-time propagator of the system can be represented as 
\begin{eqnarray}
&&\int {\cal D}\{\varepsilon _\theta \}{\cal D}\{\varepsilon _\phi \}\int 
{\cal D}\{\theta _1\}{\cal D}\{\phi _1\}\times  \nonumber \\
&&\exp \left( -\int d\tau {\cal L}\left( \theta _1,\phi _1,\varepsilon
_\theta ,\varepsilon _\phi \right) \right) .  \label{propagator}
\end{eqnarray}
By simple algebra, up to the second order approximation about $\varepsilon
_\theta $ and $\varepsilon _\phi $, we obtain 
\begin{eqnarray}
{\cal L} &=&i2S\dot{\phi}+2\left( K_1\cos ^2\theta +K_2\sin ^2\theta \sin
^2\phi \right)  \nonumber \\
&&+\left( -iS\dot{\phi}\sin \theta -K_1\sin 2\theta \right.  \nonumber \\
&&+\left. K_2\sin 2\theta \sin ^2\phi -g\mu _BSH\sin \theta \varepsilon
_\theta \right)  \nonumber \\
&&+\left( K_2\sin ^2\theta \sin 2\phi \right) \varepsilon _\phi  \nonumber \\
&&+iS\left[ \left( 1+\cos \theta \right) -\sin \theta \varepsilon _\theta
\right] \dot{\varepsilon}_\phi  \nonumber \\
&&+\left( A_{\theta \theta }\varepsilon _\theta ^2+A_{\theta \phi
}\varepsilon _\theta \varepsilon _\phi +A_{\phi \phi }\varepsilon _\phi
^2\right) ,  \label{lag}
\end{eqnarray}
where 
\begin{eqnarray}
A_{\theta \theta } &=&-i\frac S2\cos \theta \dot{\phi}+\frac J2-K_1\cos
2\theta  \nonumber \\
&&+K_2\cos 2\theta \sin ^2\phi -\frac{g\mu _BSH}2\cos \theta ,  \nonumber \\
A_{\theta \phi } &=&K_2\sin 2\theta \sin 2\phi ,  \nonumber \\
A_{\phi \phi } &=&\frac J2\sin ^2\theta +K_2\sin ^2\theta \cos 2\phi .
\label{coefficients}
\end{eqnarray}
In Eqs. (\ref{lag}) and (\ref{coefficients}), we have dropped the subscript
of $\theta _1$ and $\phi _1$ for clarity. Upon Gaussian integrating (\ref
{propagator}) over $\varepsilon _\theta $ and $\varepsilon _\phi $ one can
obtain the following effective Lagrangian 
\begin{eqnarray}
{\cal L}_{eff} &=&i2S\dot{\phi}+2\left( K_1\cos ^2\theta +K_2\sin ^2\theta
\sin ^2\phi \right) +  \nonumber \\
&&\frac{S^2}{2J}\left[ \sin ^2\theta \left( \dot{\phi}-ig\mu _BH\right) ^2+%
\dot{\theta}^2\right] .  \label{effLag}
\end{eqnarray}
Note that magnetic field enters only through the last term in Eq. (\ref
{effLag}) and has no influence on the tunneling barrier. Because of the
condition $K_1>K_2,$ the equilibrium orientations of $\vec{S}_1$ are $\left(
\theta ,\phi \right) =\left( \frac \pi 2,0\right) $ and $\left( \frac \pi 2%
,\pi \right) $ which correspond to two degenerate classical minima of the
energy, $E=0$. It is obvious from symmetry that there are two different type
instanton trajectories of opposite windings around hard anisotropy axis. We
denote them as $\pm $ instantons: 
\begin{equation}
\phi =0\longrightarrow \phi =\pm \pi /2\longrightarrow \phi =\pm n\pi \left(
n=1,3,5,...\right) .
\end{equation}

To execute the first, we should seek the classical path(or paths) $\Omega
_{cl}\left( \tau \right) =\left( \bar{\theta}\left( \tau \right) ,\bar{\phi}%
\left( \tau \right) \right) $ connecting the two minima, that minimizes the
action $S_E=\int d\tau {\cal L}_{eff}$. This path satisfies the
Euler-Lagrange equations of motion (see also Eqs. (\ref{Ea}) and (\ref{Eb})) 
\begin{equation}
\frac d{d\tau }\left( \frac{\partial {\cal L}_{eff}}{\partial \dot{\Omega}%
_{cl}\left( \tau \right) }\right) -\frac{\partial {\cal L}_{eff}}{\partial
\Omega _{cl}\left( \tau \right) }=0.  \label{eom}
\end{equation}
Substituting the effective Lagrangian into Eq. (\ref{eom}), we obtain 
\begin{eqnarray}
&&\frac d{d\tau }\left[ \frac{S^2}J\frac{d\bar{\theta}}{d\tau }\right] 
\nonumber \\
&=&\left( -2K_1+2K_2\sin ^2\bar{\phi}\right) \sin 2\bar{\theta}  \nonumber \\
&&+\frac{S^2}{2J}\sin 2\bar{\theta}\left( \frac{d\bar{\phi}}{d\tau }-ig\mu
_BH\right) ^2 \\
&&  \nonumber \\
&&\frac d{d\tau }\left[ \frac{S^2}J\sin ^2\bar{\theta}\left( \frac{d\bar{\phi%
}}{d\tau }-ig\mu _BH\right) \right]  \nonumber \\
&=&2K_2\sin ^2\bar{\theta}\sin 2\bar{\phi}.  \label{sg}
\end{eqnarray}
Consequently, a quasiclassical tunneling of $\vec{S}_1$ may occur in $xy$%
-plane $\bar{\theta}=\frac \pi 2$, and then, Eq. (\ref{sg}) reduces to
sine-Gordon equation, 
\begin{equation}
2\frac{d^2}{d\tau ^2}\bar{\phi}=\omega _1^2\sin 2\bar{\phi},
\end{equation}
or equivalently 
\begin{equation}
\frac{d\bar{\phi}}{d\tau }=\omega _1\sin \bar{\phi},
\end{equation}
where $\omega _1=\left( \frac{4JK_2}{S^2}\right) ^{1/2}$. Under the boundary
conditions in which the classical path approach the two minima as $\tau
\rightarrow \pm \infty $, we obtain an exact solution of this equation \cite
{barbara}, 
\begin{equation}
\bar{\phi}\left( \tau \right) =2\arctan \left( \exp \left( \omega _1\tau
\right) \right) .
\end{equation}
It is easily verified that $\bar{\phi}\rightarrow 0,\pi $, as $\tau
\rightarrow \pm \infty $. The corresponding classical action, i.e., the WKB
exponent in the rate of quantum tunneling at finite magnetic field, can be
evaluated by integrating the Euclidean action with above classical
trajectories, and the result is found to be 
\begin{equation}
S_{cl}^{\pm }=%
\mathop{\rm Re}%
S_{cl}\pm i%
\mathop{\rm Im}%
S_{cl},  \label{action}
\end{equation}
with 
\begin{eqnarray}
\mathop{\rm Re}%
S_{cl} &=&4S\left( \frac{K_2}J\right) ^{1/2},  \label{re} \\
\mathop{\rm Im}%
S_{cl} &=&2\pi S\left( 1-\frac H{H_a}\right) ,  \label{im}
\end{eqnarray}
where the positive and negative sign in Eq. (\ref{action}) are corresponding
to $\pm $ instantons, respectively.

It is clearly seen from Eqs. (\ref{re}) and (\ref{im}), the classical action
has two unusual features in the presence of magnetic field. First of all,
the real part of action has no dependence on the magnetic field and is
determined by material parameters of the system only. This feature is quite
different from that in ferromagnetic molecular magnets, and can be
understood easily from Eq. (\ref{effLag}), since the tunneling barrier
remains unchanged under the magnetic field. Further, as shown from Eq. (\ref
{im}), if we ignore the contribution from quantum fluctuations around the
classical paths, the ground-state tunnel splitting which is proportional to $%
\exp (-%
\mathop{\rm Re}%
S_{cl})\left| \cos (%
\mathop{\rm Im}%
S_{cl})\right| $ oscillates as the field $H$ is increased, and the tunneling
is thus quenched whenever 
\begin{equation}
H=\frac{\left( 2S-n-1/2\right) }{2S}H_a,
\end{equation}
where $n=0,1,2,...$. It is interesting to note that this result agrees well
with Eq. (\ref{effLag}) in Ref. \cite{garg93} found by Garg for
ferromagnetic molecular Fe$_8$ clusters if one makes the replacement $J=2S$
and sets $\lambda =0$.

\subsubsection{preexponential factor}

The second major step is to evaluate the preexponential factor of small
fluctuations around the classical instanton paths. We write 
\begin{eqnarray}
\theta \left( \tau \right)  &=&\bar{\theta}\left( \tau \right) +\delta
\theta \left( \tau \right) ,\qquad  \\
\phi \left( \tau \right)  &=&\bar{\phi}\left( \tau \right) +\delta \phi
\left( \tau \right) ,
\end{eqnarray}
and evaluate the action to second order in $\left( \delta \theta ,\delta
\phi \right) $. Writing $S_E=S_{cl}+\delta ^2S$, we have 
\begin{eqnarray}
\delta ^2S &=&\int d\tau \frac{S^2}{2J}\left\{ \delta \dot{\theta}^2+\left[
\left( g\mu _BSH\right) ^2+\omega _0^2-\omega _1^2\right. \right.   \nonumber
\\
&&+\left. \omega _1^2\cos 2\bar{\phi}\pm i2g\mu _BSH\omega _1\sin \bar{\phi}%
\right] \delta \theta ^2  \nonumber \\
&&+\left. \delta \dot{\phi}^2+\left( \omega _1^2\cos 2\bar{\phi}\right)
\delta \phi ^2\right\} ,  \label{d2S}
\end{eqnarray}
where $\omega _0=\left( \frac{4JK_1}{S^2}\right) ^{1/2}$. Note that the $%
\theta $ and $\phi $ fluctuations are decoupled in Eq (\ref{d2S}), and in
the $\theta $-fluctuation, an unusual term $\pm i2g\mu _BSH\omega _1\sin 
\bar{\phi}$ distinguishing $+$ instantons from $-$ instantons appears. As we
will show below, this extra term has important consequences at the high
magnetic field.

Now, the imaginary-time propagator is found to be 
\begin{equation}
\left\langle \hat{n}_f\left| exp\left[ -{\cal H}T\right] \right| \hat{n}%
_i\right\rangle =\exp \left( S_{cl}^{\pm }\right) D_{\delta \theta }^{\pm
}D_{\delta \phi }
\end{equation}
with 
\begin{eqnarray}
D_{\delta \theta }^{\pm } &=&{\cal N}_\theta \int {\cal D}\{\delta \theta
\}\times   \nonumber \\
&&\exp \left\{ -\int d\tau \left[ \frac{S^2}{2J}\delta \dot{\theta}^2+\left(
\left( g\mu _BSH\right) ^2+\omega _0^2\right. \right. \right.   \nonumber \\
&&\left. \left. \left. -\omega _1^2+\omega _1^2\cos 2\bar{\phi}\pm i2g\mu
_BSH\omega _1\sin \bar{\phi}\right) \delta \theta ^2\right] \right\} , \\
D_{\delta \phi } &=&{\cal N}_\phi \int {\cal D}\{\delta \phi \}\times  
\nonumber \\
&&\exp \left\{ -\int d\tau \frac{S^2}{2J}\left[ \delta \dot{\phi}^2+\left(
\omega _1^2\cos 2\bar{\phi}\right) \delta \phi ^2\right] \right\} ,
\end{eqnarray}
where ${\cal N}_\theta $ and ${\cal N}_\phi $ are the normalization factors.
The fluctuation determinant for $\phi $ is standard. Following the Eq.
(2.44) in Ref. \cite{garg92}, we obtain 
\begin{equation}
D_{\delta \phi }=2\omega _1\left( \frac{S^2\omega _1}{J\pi }\right)
^{1/2}=2\omega _1\left( \frac{%
\mathop{\rm Re}%
S_{cl}}{2\pi }\right) ^{1/2}.
\end{equation}
For the $\theta $-fluctuation determinant we find in the first order
perturbation theory (The detailed calculation of $D_{\delta \theta }^{\pm }$
will be reported elsewhere.), for the high magnetic field, 
\begin{equation}
D_{\delta \theta }^{\pm }=\exp \left( \frac{\eta ^{1/2}\mp i\frac \pi 2}{%
\left( 1+\frac{K_1}{K_2}\eta \right) ^{1/2}}\right) ,  \label{det}
\end{equation}
where $\eta =\frac{\omega _1}{g\mu _BSH}=\frac{H_a}H\left( \frac{K_2}J%
\right) ^{1/2}$ is used as a small parameter in the high magnetic field. It
is clearly shown in Eq. (\ref{det}), the existence of magnetic field can
bring a phase shift which approaches approximately $\frac \pi 2$ in the
high-field regime. Note that the value of phase shift agrees well with that
found by Chiolero and Loss \cite{chiolero}. On the other hand, as the field
decreases down to zero, and thus $\eta \rightarrow \infty $, the phase shift
vanishes despite the breakdown of our first order perturbation calculations
in the low-field regime.

Combing the classical action and two fluctuation determinants, we arrive at
the desired ground-state tunnel splitting, 
\begin{eqnarray}
\Delta &=&4\exp \left( \frac{\eta ^{1/2}}{\left( 1+\frac{K_1}{K_2}\eta
\right) ^{1/2}}\right) \omega _1\left( \frac{%
\mathop{\rm Re}%
S_{cl}}{2\pi }\right) ^{1/2}\times  \nonumber \\
&&\exp \left( -%
\mathop{\rm Re}%
S_{cl}\right) \left| \cos \Phi (H)\right| ,  \label{splitting}
\end{eqnarray}
where 
\begin{equation}
\Phi (H)=2\pi S\left( 1-\frac H{H_a}\right) +\frac{\frac \pi 2}{\left( 1+%
\frac{K_1}{K_2}\eta \right) ^{1/2}}.
\end{equation}

\section{Results and discussions}

The semiclassical analysis presented so far applies strictly speaking only
to a sizable number of spins with $S\gg 1$. However, as is often the case
with such methods the results are valid (at least qualitatively) even down
to a few spins of small size. This expectation is indeed confirmed by exact
diagonalization calculations which we have performed on Hamiltonian (\ref
{totHami}). Result for $S=5$, and for some typical values $k_1=0.03$ {\rm K}%
, $k_2=0.01$ {\rm K} and $j=1.0$ {\rm K} is presented in Fig. 1, the
critical field is found to be 7.44 {\rm T}. Here, the units for the energy
and magnetic field are taken to be Kelvin and Tesla, respectively. We can
see that the numerical and semiclassical approach show reasonable agreement
in the whole magnetic field regime. Since our perturbed calculation for the $%
\theta $-fluctuation determinant is only valid in the high magnetic field,
the agreement in the low-field regime is surprising in some ways.

As shown in Fig.1, the ground-state tunnel splitting vanishs at the field $%
\frac H{H_a}=1.0$. This disappearance is evident for the extra $\frac \pi 2$
phase shift, since according to Eq. (\ref{im}), there should be a peak in
usual. It is worthy noting that the extra $\frac \pi 2$ phase shift is not
limited to our biaxial symmetry antiferromagnetic molecular magnets case.
Indeed, for the strong antiferromagnetic coupling, upon Gaussian integrating
(\ref{propagator}) over the small displacements $\varepsilon _\theta $ and $%
\varepsilon _\phi $, one can obtain the effective Lagrangian in general form 
\begin{eqnarray}
{\cal L}_{eff} &=&i2S\dot{\phi}  \nonumber \\
&&+E\left( \theta ,\phi \right) +\frac{S^2}{2J}\left[ \sin ^2\theta \left( 
\dot{\phi}-ig\mu _BH\right) ^2+\dot{\theta}^2\right] ,
\end{eqnarray}
where the detailed form of $E\left( \theta ,\phi \right) $ depends on the
system investigated. Then, an extra term similar to $\pm i2g\mu _BSH\omega
_1\sin \bar{\phi}$ in Eq. (\ref{d2S}) will appear after we expand the small
fluctuations around the classical instanton paths to the second order, and
gives the $\frac \pi 2$ phase shift. Thus, we conclude that the $\frac \pi 2$
phase shift induced by $\theta $-fluctuation may be a universal quantum
behavior in all antiferromagnetic molecular magnets systems.

In the figure, another interesting feature is the peak height of the
ground-state tunnel splitting. As the magnetic field increases, the peak
height first drops significantly in the low-field regime, and then keeps
invariant up to the critical field $H_a$. This may be qualitatively
understood from Eq. (\ref{im}). Because the WKB exponent has no relevance
with the magnetic field, the most essential dependence of the ground-state
tunnel splitting on the field comes from the preexponential factor $%
p_0=4\exp \left( \frac{\eta ^{1/2}}{\left( 1+\frac{K_1}{K_2}\eta \right)
^{1/2}}\right) $ which undergoes a dramatically change only in the low-field
regime.

At the end of this section, in order to support the experimental relevance
of our results, we give some estimates for the ferric wheel, Fe$_{10}$, for
which $N=2n=10,$ $s=\frac 52,$ $S=ns=12.5$. If one takes $\frac j{g\mu _B}=4$
{\rm T}, $\frac{k_1}j=0.03$ and $\frac{k_2}j=0.01$ as the typical parameters
values \cite{chiolero}, then the simple algebra demonstrates that the
ground-state tunnel splitting has $2S=25$ oscillations of magnitude $\Delta
\approx 1.3$ {\rm K} and period $4$ {\rm T}. From Eq. (\ref{det}), the phase
shift will be visible when $\frac{K_1}{K_2}\eta \sim 3$, or $H=10$ {\rm T}.
Therefore, all quantities appear to be well within experimental reach.

\section{Conclusion}

In summary, we have investigated the MQC phenomena in biaxial symmetry
antiferromagnetic molecular magnets. Our discussion is based on the
two-sublattice model that includes anisotropy and magnetic field. On the
basis of instanton technique in the spin-coherent-state path-integral
representation, both the rigorous Wentzel-Kramers-Brillouin exponent and
preexponential factor for the ground-state tunnel splitting are obtained. We
have outlined here two prominent features in our tunneling scenario: ($i$)
In addition to the usual topological term in the classical action, a new
quantum phase arising from the quantum fluctuations around the classical
paths is found to contribute to the tunneling oscillations. This result
coincides with that reported by Chiolero and Loss \cite{chiolero}. ($ii$)
The magnetic field appears to have no influence on the tunneling barrier.
Thus the main dependence of the tunneling peak height on the field comes
from the quantum fluctuations, this leads to a sudden drop of peak's height
in the low-field regime. Both two features clearly have no analogue in the
ferromagnetic systems \cite{garg93,garg99,wernsd}. We suggest that they may
be universal behaviors in all antiferromagnetic molecular magnets.

We realize that our result is based on the instanton method which is
semiclassical in nature, i.e., valid in large spins and in continuum limit.
We perform exact diagonalization calculation to check its validity, and find
that it agrees well with the analytical result in the regime where a
comparison is possible.

Recent experiments have rekindled interest in the field of quantum tunneling
of the molecular magnets.. Most notable has been the discovery of resonant
quantum tunneling between spin states in the ferromagnetic system of spin-$%
10 $ molecules such as Mn$_{12}$Ac \cite{Mn12Ac} and Fe$_8$ \cite{Fe8}.
Since the antiferromagnetic molecular magnets are proposed as better
candidates for observing the phenomena of MQT and MQC \cite{barbara}
compared with the ferromagnetic ones, we hope that our predictions on
antiferromagnetic molecular magnets can be confirmed in experiments in the
future.

{\bf Acknowledgments}

The financial supports from NSF-China (Grant No. 19974019) and China's
``973'' program are gratefully acknowledged.

\begin{center}
{\bf Figure caption}
\end{center}

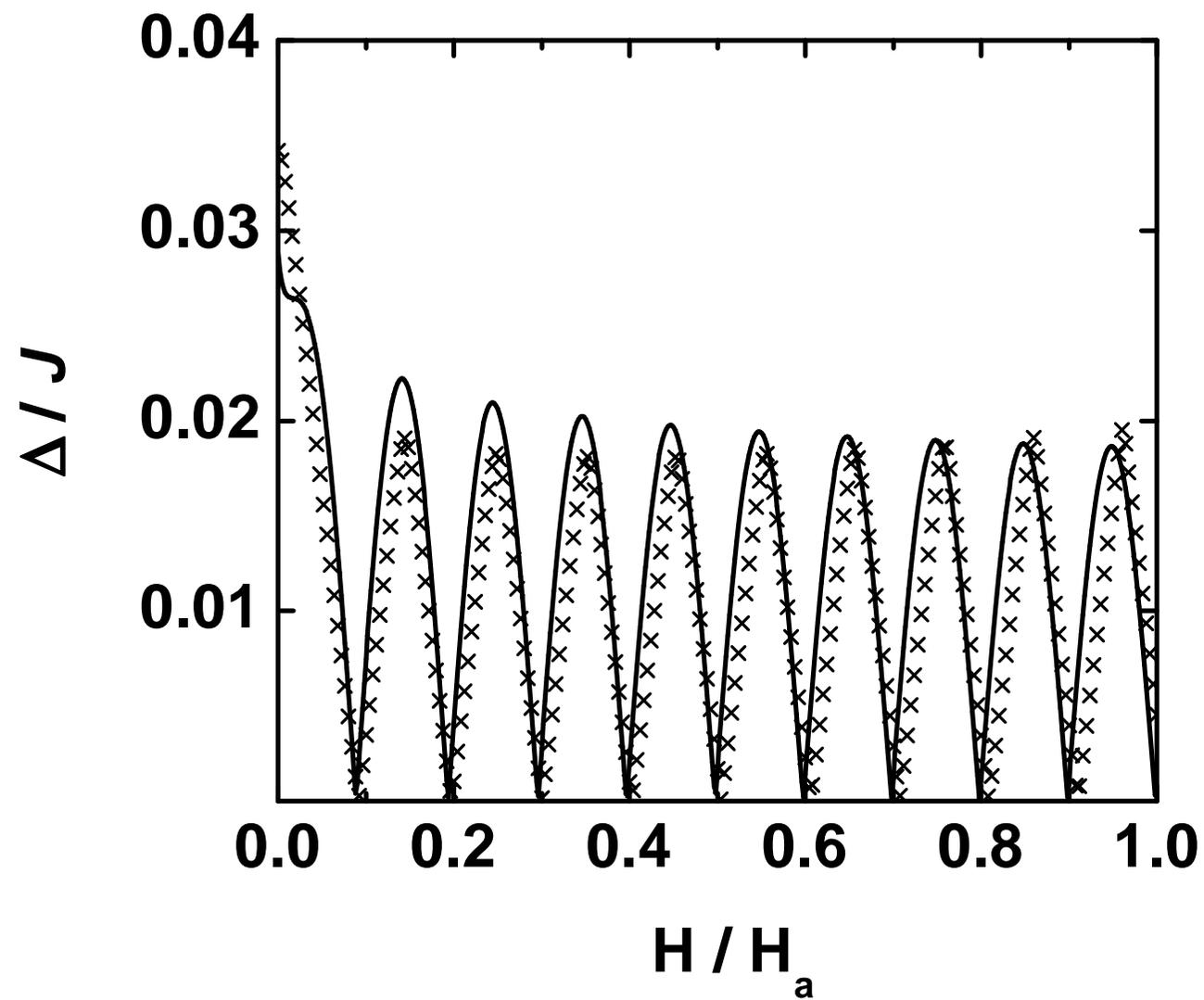
Fig.1. $\frac \Delta J$ versus $\frac H{H_a}$, with $S=5$, $k_1=0.03$ {\rm K}%
, $k_2=0.01$ {\rm K}, $j=1.0$ {\rm K} and $H_a=7.44$ {\rm T}. Here, the
units for the energy and magnetic field are taken to be Kelvin and Tesla.
The solid line and symbols represent, respectively, the ground-state tunnel
splitting predicted by the semiclassical instantons approach and that
obtained by the exac diagonalization method.

\end{document}